\documentstyle[11pt,aaspp4]{article}
\begin{document}
\def\etal{{\it et al.\/}}
\def\cf{{\it cf.\/}}
\def\ie{{\it i.e.\/}}
\def\eg{{\it e.g.\/}}

\title{A Self-Similar Solution for the Propagation of a  Relativistic 
Shock in an Exponential Atmosphere}
\author{Rosalba Perna\altaffilmark{1,2} and Mario Vietri\altaffilmark{3}}
\altaffiltext{1}
{Harvard Junior Fellow, Harvard Smithsonian Center for Astrophysics, 60 
Garden Street, Cambridge, MA, 02138}
\altaffiltext{2}{Osservatorio Astronomico di Roma, via dell'Osservatorio, 2,
I-00040, Monte Porzio Catone, Roma, Italy}
\altaffiltext{3}{Universit\`a di Roma 3, Via della Vasca Navale 84, 00147 Roma, Italy}

\begin{abstract}
We derive a fully relativistic, self--similar solution to describe the
propagation of a shock along an exponentially decreasing atmosphere,
in the limit of very large Lorentz factor. We solve the problem in
planar symmetry and compute the acceleration of the shock in terms of
the density gradient crossed during its evolution.  We apply our
solution to the acceleration of shocks within the atmosphere of a
HyperNova, and show that velocities consistent with the requirements of
GRB models can be achieved with exponential atmospheres spanning a
wide density range.
\end{abstract}


\section{Introduction}

In a SuperNova (SN) explosion, the shock wave must ultimately emerge
from the body of the star, and begin to propagate down the exponential
density gradient of the stellar atmosphere. In Newtonian fluid
dynamics, the propagation of such a shock is described by a
self--similar solution (Raizer 1964; Grover and Hardy 1966; Hayes
1968). This self--similar solution is unavoidable (Raizer 1964). In
fact, it turns out that, as the shock accelerates, a sonic point is
formed, separating matter located immediately behind the shock from
the flow's initial conditions; this prevents pressure waves ({\it
i.e.}, causal information) to reach post--shock material from the the
area where the flow's initial conditions are set. Thus all flows will
converge to the same solution, irrespective of their initial
conditions. The lack of dependence upon initial conditions severely
constrains the ensuing flow, by restricting the number of parameters
upon which the shock evolution may depend; in fact, this constraint is
so strong that only a single (self--similar) solution exists.

In the Newtonian solution, it is found that the shock
velocity increases exponentially, in such a way that the shock reaches
spatial infinity in a finite time.  This is clearly impossible when
account is taken of Special Relativity, and therefore the Newtonian
solution must break down at high shock velocities, as expected.  The
question of precisely how the shock evolves in relativistic conditions
has not been investigated so far.

This problem has acquired a new urgency within the HyperNova model 
(Paczy\'nski 1998; MacFadyen \& Woosley 1999) for Gamma 
Ray Bursts (GRBs). In fact, Meszaros and Rees (2001) have shown that, despite 
the large assumed energy release of the central engine, the outwardly moving 
shock in the large star hypothesized to give rise to the GRB can only reach a 
Lorentz factor of $\Gamma_i \approx 10$ at the end of the H--envelope, and they 
had to invoke (without explicit computations) shock acceleration down the star 
exponential atmosphere 
to reach the required Lorentz factors of $\Gamma_f \approx 100-300$. 
The existence of a self-similar temporal structure in GRBs is also
suggested by an analysis of their power spectra (Beloborodov, Stern,
\& Svensson 2000; see also Sivron 1998).

It is the purpose of this paper to derive a fully (Special) relativistic 
self--similar solution for the problem of shock propagation in an exponential
atmosphere (Section 2), and to discuss its use in the HyperNova model for
GRBs (Section 3).

\section{Relativistic self--similar flow}

We shall assume that a cold ($T = 0$) material is stratified with 
density distribution $\rho = \rho_\circ \exp(- k x)$; the shock 
is supposed to move toward $x = +\infty$, thus with positive velocity $v > 0$.
The symmetry is assumed to be planar, since both the length scale of
stellar atmospheres and the total extension of the atmosphere are much 
smaller than the stellar radius. In the Newtonian analog, the shock speed is set 
by a purely dimensional argument (e.g. Chevalier 1990): $V = \alpha/ k t$, 
with $\alpha$ an adimensional constant
to be determined. In the special relativistic problem, dimensional arguments
alone fail because the presence of the light speed $c$ allows the construction 
of a new adimensional quantity ($k c t$), thusly spoiling the above argument.
We can however recover from this impasse by appealing to both dimensional and
covariance arguments. The time--like part of the shock four--speed $U_\mu$ is 
of course determined by the identity $U_\mu U^\mu = -1$, while the space--like
part must be built in covariant fashion from the available quantities. Defining
a four vector $k_\mu = (0, k, 0, 0)$, we can then write for the spatial part of
$U$
\begin{equation}
\label{propertime}
U^a = \frac{d\!X^a}{d\!s} = \frac{\alpha k^a}{k_\mu k^\mu s}
\end{equation}
where $k_\mu k^\mu = k^2 > 0$. 
This is the only sensible solution to our dimensional/covariant
problem. We see from this that, as long as we use proper time, the
structure of the problem is identical to that of the Newtonian
analog. In particular, the shock reaches spatial infinity within a
finite proper time. We may choose $s=0$ for the moment when this occurs, so
that the flow is restricted to $s<0$.  Incidentally,
note that this implies $\alpha < 0$. Physically, this makes perfect
sense: as the shock accelerates, its proper time is contracted by its
Lorentz factor ($\Gamma$) with respect to the fluid time. 
The shock speed in terms of the fluid time can be obtained by remembering that
$d\!X/d\!s = v \Gamma$, and substituing in the above equation one
finds ($c=1$ from now on)
\begin{equation}
s = \frac{\alpha}{k v \Gamma}\;.
\end{equation}
Taking the time derivative of both sides we obtain
\begin{equation}
\frac{k d\!t}{\alpha} = \Gamma \left(\frac{d}{d\!v}\frac{1}{v \Gamma}\right)
 {d\!v}
\end{equation}
which can be immediately integrated to yield
\begin{equation}
\frac{k t}{-\alpha} = \log\left(\frac{1+v}{1-v}\right)^{1/2}
- \frac{1}{v} + \mbox{constant}\;.
\end{equation}
The low speed limit $v \ll 1$ is $k t/\alpha \approx 1/v$, in agreement with the
Newtonian solution. The hyperrelativistic limit ($v\rightarrow 1$) is
\begin{equation}
\label{gammat}
\frac{k (t-t_i)}{-\alpha} \approx \log\Gamma/\Gamma_i
\end{equation}
where we have introduced an initial shock Lorentz factor $\Gamma_i$ at time 
$t_i$, for ease of use in the future. Since the shock moves essentially at
speed $1$, we can rewrite the lhs of the above equation as $\log (\rho/\rho_i)^
{1/\alpha}$, from which we see that the initial and the final (i.e., when the shock
leaves the exponential atmosphere) shock Lorentz factors are related by
\begin{equation}
\label{important}
\frac{\Gamma_f}{\Gamma_i} = \left(\frac{\rho_f}{\rho_i}\right)^{1/\alpha}\;.
\end{equation}

The above clearly shows that all we have left to do is to determine 
the value of the parameter $\alpha$. To this purpose,
we must consider the fluid equations for the post--shock material, which we 
have found it convenient to take in the form given by Blandford and McKee 
(1976, their Eqs. 14-15)
\begin{equation}
\label{bmck1}
\frac{d}{d\!t} (e \gamma^4) = \gamma^2\frac{\partial\!e}{\partial\!t}
\end{equation}
\begin{equation}
\label{bmck2}
\frac{d}{d\!t} \log (e^3 \gamma^4) = - 4 \frac{\partial\!v}{\partial\!x}
\end{equation}
\begin{equation}
\frac{\partial\!n}{\partial\!t} + \frac{\partial}{\partial\!x} (v n) = 0
\label{bmck3}
\end{equation}
where $d/d\!t = \partial/\partial\!t + v\partial/\partial\!x$ is the
convective derivative, and we have dropped curvature terms (which are contained
in the Blandford and McKee equations), to keep with our planar symmetry
approach. Here $e$ is the local energy density, 
$\gamma$ is the local fluid Lorentz factor, as seen from the
reference frame of the unshocked material (to be distinguished from $\Gamma$,
the shock Lorentz factor in the same reference frame), and $n$ is the baryon 
number density always in the reference frame of the  unshocked fluid. The above 
equations assume an hyperrelativistic equation of state for the post--shock
material of the form $p = e/3$, which is correct in the limit $\Gamma
\rightarrow \infty$.  Indeed, for the large shock Lorentz factors 
appropriate to this problem ($\Gamma\ga 10$) Heavens 
\& Drury (1988; see also Iwamoto 1989) showed 
that the adiabatic index of the flow is within $1\%$ of its asymptotic
relativistic value of $4/3$, irrespective of the upstream 
temperature and plasma composition
(see especially their Fig.2, and Eqs. 25-26).
It should be remarked here that Eq. (\ref{important}) remains 
valid whatever we assume for the equation of state, but the precise value 
of $\alpha$ depends instead on the particular choice of the equation of state. 

The boundary conditions for this problem are provided by Taub's jump 
conditions (Taub 1948), which again we take in the hyperrelativistic limit given by 
Blandford and McKee (1976):
\begin{equation}
\label{taub} 
e_2 = 2\Gamma^2 w_1 \;,\;
\gamma_2^2 = \frac{1}{2}\Gamma^2\;,\;
n_2 = 2 \Gamma^2 n_1\;\;.
\end{equation}

Here the subscripts $1,2$ refer to pre-- and post-- shock quantities, 
respectively; $n$ and $\gamma$ are always defined in the unshocked fluid 
frame, and $e$ in the comoving frame.  
The enthalpy of the unshocked material, $w_1$, equals $e_1$ since
this material is assumed to be cold. In our problem, $e_1$ is not a constant, because the 
atmosphere is stratified. We have
\begin{equation}
w_1 = e_1 = \rho_1 = \rho_\circ \exp(-k X) = \rho_\circ (\Gamma/\Gamma_i)^\alpha
\end{equation}
so that, ultimately,
\begin{equation}
\label{taube}
e_2 = 2 \left(\frac{\rho_\circ}{\Gamma_i^\alpha}\right) \Gamma^{2+\alpha} \equiv 
2 q_\circ \Gamma^{2+\alpha}
\end{equation}
which is the form we shall use in the following. An identical 
argument shows that
\begin{equation}
\label{taubn} 
n_2 = 2 \frac{n_\circ}{\Gamma_i^\alpha} \Gamma^{2+\alpha} \equiv 2 z_\circ 
\Gamma^{2+\alpha}\;,
\end{equation}
where $n_\circ\equiv \rho_\circ/m$.

In order to search for a self--similar solution, we need to assume a form
for the similarity variable. In the Newtonian analog, this is clearly $\xi =
k(x-X)$, where $X$ is the instantaneous shock location. In this problem, we
take
\begin{equation}
\label{scalingvariable}
\xi \equiv k (x-X) \Gamma^2\;.
\end{equation}
The rationale for this is that, since post--shock material has a Lorentz factor
which is $\sqrt{2}$ times smaller than the shock's, the post--shock material
falls behind the shock by an amount $\propto1/\Gamma^2$ as the shock covers an
exponential length. Please notice that, with our notation, $\xi < 0$ for the shocked
material. The post--shock quantities $\gamma^2$, $e$ and $n$ can then be
taken of the form
\begin{equation}
\label{scaledquantities}
\gamma^2 = g(\xi) \Gamma^2 \;,\; e = q_\circ R(\xi) \Gamma^{2+\alpha} \;,\;
n = z_\circ \Gamma^{2+\alpha} N(\xi)
\end{equation}
with boundary conditions given by Eqs. (\ref{taub}) (the second one), 
(\ref{taube}), (\ref{taubn}) as
\begin{equation}
g(0) = \frac{1}{2}\;,\; 
R(0) = 2 \;,\; 
N(0) = 2\;.
\end{equation}

Substitution of Eqs. (\ref{scalingvariable}), (\ref{scaledquantities}) into Eqs.
(\ref{bmck1}), (\ref{bmck2}) and (\ref{bmck3}) yields, after some algebra
 (the dot indicates derivation with respect to $\xi$)
\begin{equation}
\label{grand1}
\frac{\dot{R}}{R} = \frac{-g(2/\alpha +1)(1-4\xi/\alpha + 1/g)+(10/\alpha+3)}
{3/2-3/2g -6\xi/\alpha + (1-g/2+2\xi g/\alpha)(1-4\xi/\alpha +1/g)}
\end{equation}
\begin{equation}
\label{grand2}
\frac{\dot{g}}{g^2} = \frac{\dot{R}}{R}(1/g-1/2+2 \xi/\alpha) + (2/\alpha+1)
\end{equation}
\begin{equation}
\label{grand3}
\frac{\dot{N}}{N} = \frac{2 g \frac{2+\alpha}{\alpha} - \frac{\dot{g}}{g}}{g(1-4
\xi/\alpha) -1}\;.
\end{equation}
From these, we see what fixes $\alpha$: the denominator of the rhs of Eq. 
(\ref{grand1}) goes to zero at a critical point $\xi_c$ and, unless the 
numerator simultaneously does the same 
(which will only occur for a special value of $\alpha$), a non--integrable
singularity will ensue. The same phenomenon occurs in the Newtonian analog, 
where it has been shown that this critical point is a sonic point. 
It is exactly because there is a sonic point that a self--similar solution
can develop: in fact, material between the shock and the sonic point
is not in causal contact with post--sonic point material, and thus 
its properties cannot be determined by the problem's initial conditions.

To keep the problem non--singular, we impose that the  
numerator and denominator of Eq. (\ref{grand1}) vanish simultaneously, obtaining
\begin{equation}
g(\xi_c)(1-4\xi_c/\alpha) = \frac{8+2\alpha}{2+\alpha}\;
\label{sol1}
\end{equation}
from the numerator, and
\begin{equation} 
g(\xi_c)(1-4\xi_c/\alpha) = \frac{4 \pm \sqrt{12}}{2}\;
\label{sol2}
\end{equation}
for the denominator. The conditions above lead to 
two solutions for $\alpha$, one positive and another negative. 
The positive solution must be discarded because 
otherwise the shock would not reach spatial infinity for $s\rightarrow 0$ (Eq.
\ref{propertime}). We are therefore left with only one sensible solution,
\begin{equation}
\alpha = \frac{-12 - \sqrt{192}}{6} \approx -4.309401\;.
\label{finalalfa}
\end{equation}
The location of the sonic point, $\xi_c$, where both the numerator and
the denominator vanish, is determined through numerical integration
of Eqs. (\ref{grand1}), (\ref{grand2}) and (\ref{grand3}).     
This yields  $\xi_c \approx -0.46$. Also note that
$g(\xi)(1-4\xi/\alpha)\le g(0) = 0.5$, and therefore the
denominator of Eq. (\ref{grand3}) is always well-behaved.  
The full numerical solution 
for $g(\xi)$, $R(\xi)$ and $N(\xi)$ is illustrated in Fig. 1.

\section{Application to GRBs}

It can be seen from Eq. (\ref{important}) that the necessary asymptotic Lorentz
factor required for a proper modelling of the properties of GRBs,
i.e. $\Gamma_f \ga 150$ (remember that the Lorentz factor of the matter 
is only $1/\sqrt{2}$ that of the shock) can be reached by crossing a 
modest factor of $10^5$ density decrease in the exponential atmosphere
(assuming $\Gamma_i\sim 10$). 
Yet, the total energy of material
moving at these large Lorentz factors is only modest. In fact, let us compute
the distribution of kinetic energy with 
baryon number. Consider a cylindrical fluid element with surface
area $A$ (with the normal along the direction of fluid motion $x$), and length
$d\!x$ along the direction $x$. This element will have a bulk kinetic energy
of $d\!E = A m c^2 n(x) \gamma(x) d\!x = A \rho_\circ c^2 \Gamma^{1+\alpha} 
\Gamma_i^{-\alpha} N(\xi) g^{1/2}(\xi) k^{-1}d\!\xi$. 
What we want to know is the energy distribution 
at the moment in which the shock gets out of the exponential
atmosphere. Let us indicate with $\Gamma_f$ the shock Lorentz factor
at that moment. Then
\begin{equation}
\label{dist} 
\gamma \frac{d\!E}{d\!\gamma} = 
2 A \rho_i c^2 k^{-1} \Gamma_i^{-\alpha}\Gamma_f^{1+\alpha}\frac{N(\xi) g^{3/2}(\xi)}{\dot{g}}
\end{equation}
which can be coupled to the solution $g = g(\xi)$ to yield a parametric 
representation of $\gamma d\!E/d\!\gamma$, the distribution of kinetic energy
with respect to the final Lorentz factor. The adimensional part of the
function in the previous equation is plotted in Fig. 2. The numerical factor can be estimated 
using Eq. (\ref{important}):
\begin{equation}
\gamma \frac{d\!E}{d\!\gamma} = 10^{48}\; {\rm erg} 
\left(\frac{r_H}{10^{13} \;{\rm cm}}\right)^2
\left(\frac{k^{-1}}{10^{12}\; {\rm cm}}\right)\;
\left(\frac{\rho_i}{10^{-9}\; \; {\rm cm}^{-3}}\right)\;
\left(\frac{10}{\Gamma_i}\right)^\alpha
\left(\frac{\Gamma_f}{150}\right)^{\alpha+1}
\frac{2 N(\xi) g^{3/2}(\xi)}{\dot{g}}\;.
\end{equation} 
In this expression, the values for $r_H$ (the edge of the H--shell), 
$\rho_i$ (the matter density at the end of the H--shell, and thus 
presumably at the beginning of the exponential atmosphere), and 
$\Gamma_i$ (the shock Lorentz factor at the end of the H--shell $=$
beginning of the exponential atmosphere) are taken from Meszaros and 
Rees (2001). Taking the adimensional factor from Fig.2, we see that
the kinetic energy falls short of the average isotropic burst energy 
(Schmidt 1999) by about three orders of magnitude 
(note that here what we are computing is the isotropic energy).

Before showing how to come out of this impasse, we need to consider which 
part of Fig. 2 can actually be obtained in a realistic model. The reason for 
this limitation comes from the fact that the idealized model presented here 
has been propagating down an exponential atmosphere forever, while in a
realistic model only a finite amount of matter can have converged onto this
self--similar solution, given the finite dimension of the star.
For this reason, we have plotted in Fig. 2 three ticks,
which correspond to the present position of matter which was located, before 
being reached by the shock, at $5,10,15$ exponential scale--lengths from the
shock's present location. If we think that the initial atmosphere extends for
$5,10,15$ exponential scale--lengths, then we can believe the part of Fig. 2
located to the right of their respective tickmarks. Leftward of them, the 
true, physical solution will depart from the one shown here, and the total
kinetic energy at the corresponding values of Lorentz factor will be much
smaller than the values plotted (obviously, since the physical solution has
finite mass and energy, which is not true for the idealized one). 

Let us now suppose that the putative burst progenitor possesses a wide exponential 
atmosphere, spanning some $\ga 9$ orders of magnitude in density. Then, 
for an initial Lorentz factor $\Gamma_i\approx 10$, the
shock Lorentz factor at the end of the atmosphere is $\approx 1000$ 
(Eq. \ref{important}); once again, the overall energy of matter moving 
at this speed would be modest ($\approx 10^{46}$ erg), but Fig. 2 shows that
most of the energy (before the cutoff implied by the finite extent of the
atmosphere) will come out at a Lorentz factor $\approx 0.2$ of the shock's
factor, {\it i.e.} at $\Gamma_{matter} \approx 200$. With the values above, the
total energy then amounts to $\approx 6\times 10^{52}$ erg, in reasonable
agreement with Schmidt's (1999) isotropic estimates\footnote{
Spectral properties (i.e. line production) would be similar to those
predicted by Kallaman, Meszaros \& Rees (2001).}. 
It should be noticed
that inspection of the numerical solution for large values of the
distance parameter $\xi$ shows the material to be cold ($e \propto 
n$), so that there will be negligible further acceleration of these
slower shells by $p d\!V$ work, and $\Gamma_{\rm matter} \approx 200$ remains a
good estimate of the coasting Lorentz factor. 

To summarize, the application of our solution to shock
acceleration in the atmosphere of a massive star has shown that, in
atmospheres with small density range, the amount of energy
carried by material accelerated to the typical GRB Lorentz factors
falls short of the GRB required energetics. However, for stars with
a wide density range in their atmospheres ($\ga 10^9$ orders of
magnitude), a sufficient quantity of energy is carried by later
shells of material moving at the typical GRB Lorentz factors.
In this model, the early emission would be dominated by an ultra--hard
component, due to the very large--$\Gamma$ shells slowing down in the ISM. This is
at least qualitatively 
consistent with virtually all bursts studied in some detail by BeppoSAX (Frontera
{\it et al.}, 2000, see especially their Fig. 2), which exhibit, in their first 
few seconds only, spectra 
peaking beyond the instrument's (GRBM) observing limit of $700$ keV.

\acknowledgements RP thanks the Osservatorio Astronomico di Roma for
its kind hospitality during the time that this work was carried out.

\begin{figure}
\centerline{\epsfysize=5.7in\epsffile{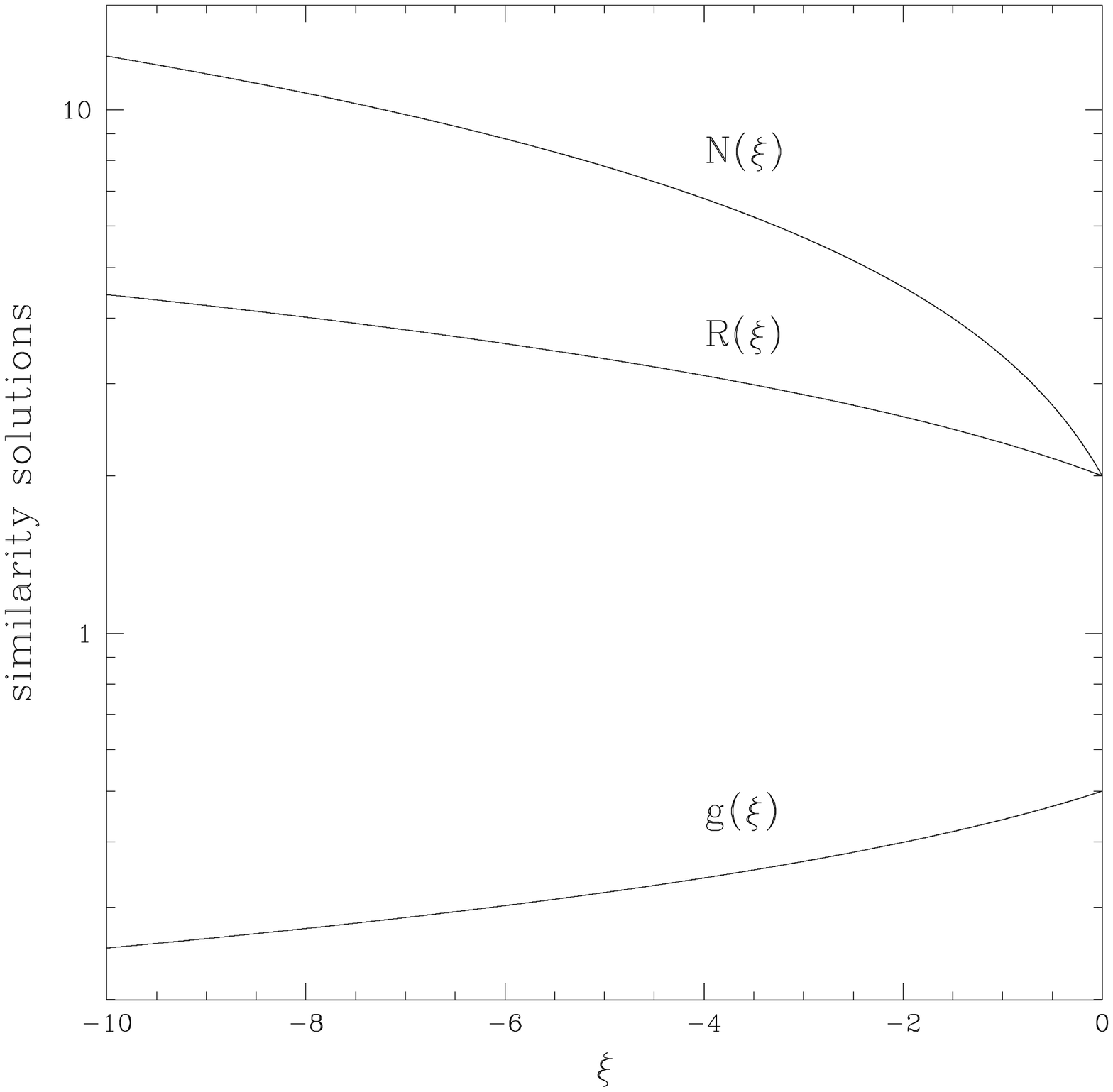}}
\caption[]{Numerical solution of Eq. (\ref{grand1}), (\ref{grand2}),
and (\ref{grand3}) with $\alpha$ 
given by Eq. (\ref{finalalfa}).}
\label{Figure 1}
\end{figure}

\begin{figure}
\centerline{\epsfysize=5.7in\epsffile{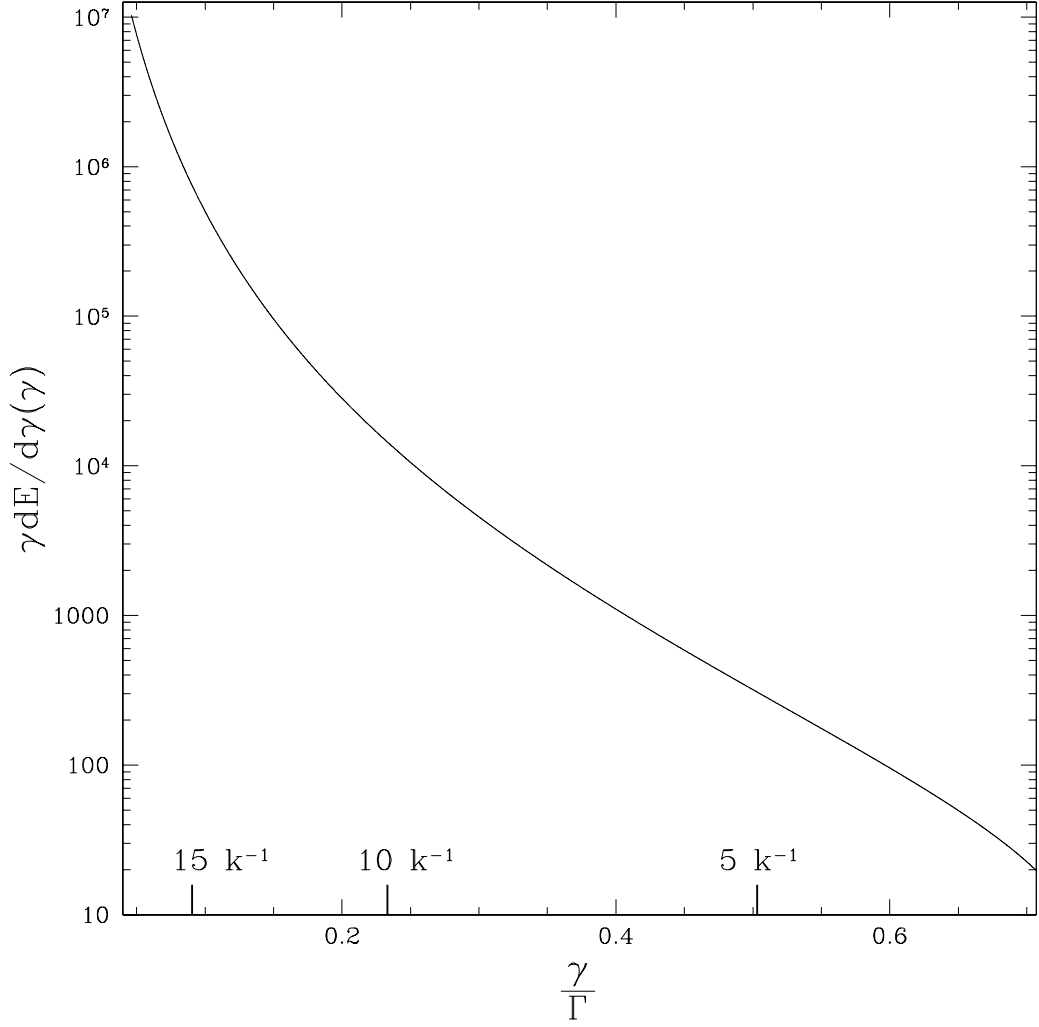}}
\caption[]{The quantity $2 N(\xi) g(\xi)^{3/2}/\dot{g}$ (the adimensional 
part of $\gamma d\!E/d\!\gamma$, Eq. \ref{dist}),
plotted as a function of $g(\xi)^{1/2}$. The ticks in boldface
indicate the present position of matter which, before being shocked,
was located at 5, 10 and 15 exponential scale lengths $k^{-1}$. } 
\label{Figure 2}
\end{figure}

\end{document}